\documentclass[pdflatex,sn-mathphys-num,iicol]{sn-jnl}

\usepackage{lmodern}
\usepackage[utf8]{inputenc}
\usepackage{graphicx}%
\usepackage{multirow}%
\usepackage{amsmath,amssymb,amsfonts}%
\usepackage{amsthm}%
\usepackage{mathrsfs}%
\usepackage[title]{appendix}%
\usepackage{xcolor}%
\usepackage{textcomp}%
\usepackage{manyfoot}%
\usepackage{booktabs}%
\usepackage{algorithm}%
\usepackage{algorithmicx}%
\usepackage{algpseudocode}%
\usepackage{listings}%

\usepackage{hyperref}
\hypersetup{
    colorlinks=true,
    linkcolor=blue,
    citecolor=blue,
    urlcolor=blue
}

\begin{document}

\title[Guided rewiring of social networks]{\textbf{Guided rewiring of social networks reduces polarization and accelerates collective action}}


\author*[1]{\fnm{Jordan P.} \sur{Everall}\,\orcid{https://orcid.org/0000-0002-2441-166X}}\email{jordan.everall@uni-graz.at}

\author[1]{\fnm{Lilli} \sur{Frei}}

\author[1,2]{\fnm{Andrew K.} \sur{Ringsmuth}\,\orcid{https://orcid.org/0000-0002-2984-8860}}

\affil*[1]{\orgname{Wegener Center for Climate and Global Change, University of Graz}, \country{Austria}}

\affil[2]{\orgname{Complexity Science Hub}, \city{Vienna}, \country{Austria}}


\abstract{Global social and ecological challenges represent collective action problems requiring rapid and sufficient cooperation with pro-mitigation norms. Sociopolitical polarization hinders such cooperation. Prior agent-based models showed polarization emerges naturally in structured social networks and polarized cluster dissolution rate limits consensus formation rate. Here we study how guided link rewiring affects depolarization dynamics across synthetic and empirical (Facebook, Twitter) network topologies. We compare heuristic rewiring algorithms representing random meetings, mutual acquaintance introductions, and community bridging, alongside topology-based link recommender algorithms (Who to Follow and node2vec). Our heuristic algorithms all outperform Who to Follow in generating cooperative consensus. Homophilic rewiring generates cooperative consensus when agents can easily change opinions. However, heterophilic rewiring achieves this over broader conditions and can accelerate cooperative consensus formation by $\approx20\%$, including where up to $33\%$ of the population experiences backfiring interactions. Heterophilic rewiring also vastly outperforms topology-based recommender algorithms. Random rewiring performed consistently well, achieving higher steady-state cooperation than seven out of eight more complex algorithms. Large disparities in steady-state cooperation for topology-based recommender systems highlight their volatility across network structures. Overall, our work reveals a subtle interplay between topology, rewiring algorithm and social depolarization, suggesting strong potential for carefully redesigning social networking technologies for social good.}


\maketitle

\section{Introduction}\label{sec:introduction}

Although the urgency of global socio-ecological challenges such as climate action and pandemic containment has become clearer \citep{ipcc_climate_2022, ringsmuth_2022}, governments struggle to enforce mitigation policies that facilitate collective action \citep{ringsmuth_2022,vasconcelos_segregation_2021,pawloff_contrarians_2014}. Political and social polarization around such global challenges is high \citep{carothers_democracies_2019,gladston_social_2019,mccarty_polarization_2019} and significantly inhibits collective action \citep{vasconcelos_segregation_2021,carothers_democracies_2019,axelrod_preventing_2021}, which suggests depolarization could strengthen functional policy-making and democracy \citep{vasconcelos_segregation_2021,pawloff_contrarians_2014,axelrod_preventing_2021}. A polity is polarized when dominated by two opposite or contradictory tendencies lacking commonality \citep{stavrakakis_paradoxes_2018}. Since individuals' opinions are strongly shaped by social context \citep{janssen_evolution_2008,gowdy_behavioral_2008,tavoni_survival_2012}, polarization is catalyzed by social structures that exclude external information and thus reinforce ideological separation \citep{gladston_social_2019,williams_network_2015,nguyen_echo_2020}.

For example, recommender systems frequently used by very large online platforms (VLOPs) like X (formerly Twitter), Meta, and TikTok often rewire links between users to maximize business performance metrics coupled with engagement, virality or advertisement exposure \cite{pappalardo_survey_2024}.  These objectives are frequently incongruent with socially beneficial network states, leading to extreme homophily \cite{stoica_algorithmic_2018,starniniEmergenceMetapopulationsEcho2016}, rich-get-richer phenomena \cite{merton_matthew_1988, su_effect_2016} and other network inequalities \cite{cinus_effect_2022}.  Asymmetric (directed) networks with influential users distributing content to many followers may exacerbate these effects.  Besides network effects, recommender systems work on categorical attributes (user traits), which together with the former produce recommendations that may degrade social cohesion and increase polarization \cite{santos_link_2021, lasser_designing_2025}. These dynamics are highly influential, with $68.7\%$ of the global population using social media \citep{noauthor_internet_nodate}. This problem, and the need for alternative rewiring of digital commons, are increasingly recognised in the legal sphere, with legislative responses such as the EU Digital Services Act (DSA) recently coming into effect.

Evidence suggests that rewiring social links (link rewiring) to increase interaction between disagreeing agents can accelerate collective action and reduce polarization under some conditions \cite{pappalardo_survey_2024}. However, the efficacy of link rewiring varies substantially between contexts. Modelling studies show that depolarization rates depend critically on the frequency of interactions between different opinion clusters and the response dynamics that govern opinion exchange \citep{axelrod_preventing_2021,fotouhi_conjoining_2018,andersson_dynamics_2021}. Some research \cite{borges_how_2024}, for example, suggests that a mix of homophily and heterophily are required for effective depolarization. The \textit{backfire effect}, in which interactions between strongly disagreeing individuals may reinforce their polarization, adds further complexity. This has been demonstrated in both online interactions \cite{bail_exposure_2018} and behavioural experiments \citep{santoroPromisePitfallsCrosspartisan2022}. Estimating its prevalence across different contexts, however, remains difficult \cite{wood_elusive_2019}.

Despite extensive research on recommender systems' polarizing effects \citep{pappalardo_survey_2024}, critical gaps remain in understanding how to harness these algorithms to instead promote depolarization and prosocial collective action. First, while simulation and empirical studies have analysed conventional recommender algorithms' impact on network-inequalities \cite{ferrara_link_2022, espin-noboa_inequality_2022}, and in other cases explored idealized rewiring strategies with opinion dynamics \citep{santos_link_2021, borges_how_2024, sasahara_social_2021}, few have integrated conventional recommender algorithms with opinion dynamics frameworks. This limits our understanding of how real adaptive network dynamics co-evolves with opinion dynamics. Second, the interplay between link rewiring and the backfire effect is not yet understood for collective action dilemmas \cite{pappalardo_survey_2024}. 

To address these gaps, we systematically evaluate how conventional recommender algorithms, and alternatives based on simple heuristics, affect the emergence of cooperative consensus across diverse network structures (empirical and synthetic). To do so, we extend an agent-based model (ABM) formulated by Andersson et al. \cite{andersson_dynamics_2021}, who investigated how polarization affects collective action in the context of large-scale societal challenges. Crucially, the model avoids assuming `bounded confidence' - that agents interact only if their opinions are sufficiently similar - because it is rational for agents to overcome bounded confidence when they understand the urgency of the societal challenge. Additionally, letting agents interact without opinion bounds highlights the importance of network structure, which bounds interactions topologically. Andersson et al \citep{andersson_dynamics_2021} analysed the dynamics on static networks. We extend this to dynamic networks by comparing how different link rewiring strategies modify the dynamics observed by Andersson et al. We compare all structured rewiring algorithms against two baselines: the static network (no rewiring) of the original model and also purely random rewiring.

\begin{figure*}[!t]
    \centering
    \includegraphics[width=\textwidth]{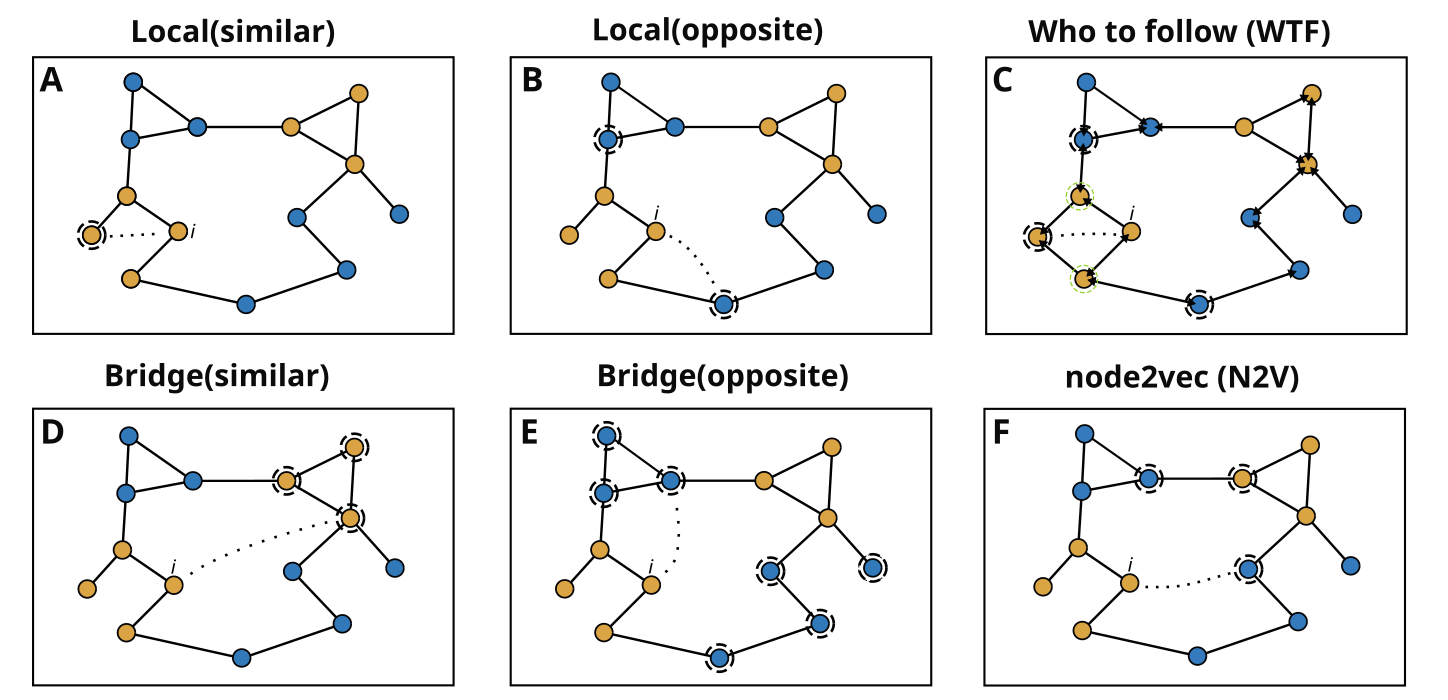}
    \caption{\textbf{Rewiring algorithms.} The colour of a node shows the sign of its opinion (blue = cooperator, orange = defector). We focus on individual \(i\) and mark available rewiring targets with dashed circles. \textbf{A} and \textbf{B} show local rewiring, in which \(i\) can choose among agents within 2 network steps (friends of friends). Dashed lines show an exemplary new link. In \textbf{A} (\textbf{B}) \(i\) only links to an agent whose opinion is similarly (oppositely) signed. In \textbf{D} and \textbf{E}, \(i\) can establish links to agents outside its own topological cluster and the opinion constraint (similar, opposite) is the same as above. \textbf{C} illustrates the Who to Follow algorithm, which calculates a circle of trust (COT, green dashed lines) for $i$ and suggests a connection to an agent commonly followed by the COT. \textbf{F} shows node2vec, which calculates and compares embeddings for all nodes, then suggests that $i$ should link to the agent with the highest structural similarity to itself. On directed networks, we assume an agent $i$ chooses others to follow by selecting from outgoing links.}
    \label{algos}
\end{figure*}

 Our model simulates the spread of cooperation with a pro-social behavioural norm on dynamic social networks. Following \cite{andersson_dynamics_2021}, we adopt evolutionary game-theoretic nomenclature (cooperation/defection relative to the norm) to reflect the strategic nature of collective action. As it is difficult to infer offline behaviour from online interactions, we proxy behavioural intentions with continuous opinions evolving through social influence \citep{ajzenTheoryPlannedBehavior1991}. An opinion takes a real number between 1 and -1, respectively representing intention to fully cooperate and intention to fully defect. We refer to agents with positive opinions ($a_i>0$) as cooperators, and those with negative opinions ($a_i<0$) as defectors. Population-wide cooperation (cooperative consensus) promotes collective action on shared challenges.

We studied the spread of cooperation with two widely-used \cite{espin-noboa_inequality_2022} link-recommender algorithms - \textit{Who to Follow}, (WTF) and \textit{node2vec} (N2V) - alongside two heuristic-based algorithms, which link agents to others in either their local network community (`local rewiring') or topologically distant communities (`bridge rewiring'), based only on topology and independent of opinions. To isolate how opinions, in contrast to topology, affect the dynamics, we introduce two variations of local and bridge rewiring which constrain the formation of new links to agents with the same or different opinion. We tested these algorithms (Fig. \ref{algos}) across four different types of network topology, two synthetic and two empirical (Facebook and Twitter). 

The conventional, topology-based recommender algorithms used here (WTF, N2V) should be distinguished from the more complex recommender systems that combine complex user data, preferences, and topological information to make link recommendations \citep{tornberg_simulating_2023}.  Although algorithms like WTF and N2V may form part of these hybrid systems, in this work we focus solely on the topological aspect, i.e., where N2V and WTF make opinion-agnostic recommendations. We refer to WTF and N2V as recommender algorithms  to reflect their real-world origins, while the term rewiring algorithms is used when collectively referring to all algorithms, i.e., both recommender-based and heuristic. In addition to different rewiring algorithms, we also investigated how different levels of the backfire effect influenced cooperation. Our findings add to the evidence that promoting connections between polarized communities could accelerate collective action on urgent global challenges, even accounting for the effect of backfiring interactions.
\begin{figure*}[!t]
    \centering
    \includegraphics[width=\textwidth]{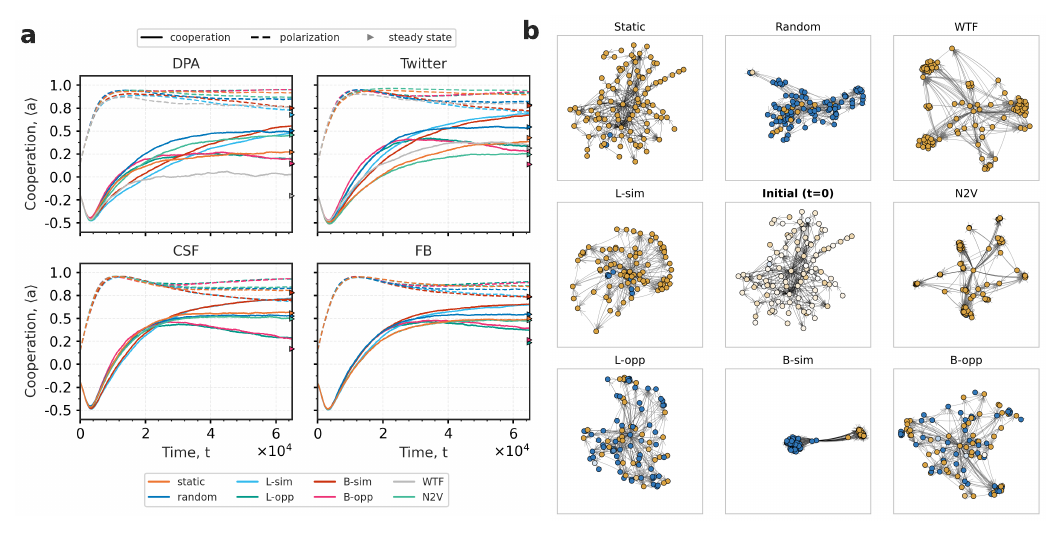}
    
    \caption{{Comparing the evolution of network structure and cooperation across all rewiring scenarios.} (a) A dynamic network topology does not guarantee higher cooperation or lower polarization than a  static baseline (orange line). Of those that do, $x(\text{similar})$ and random rewiring algorithms consistently support the emergence of positive cooperation in the steady state (solid arrows). (b) Representative ($N=100)$ network states for selected rewiring algorithms, given a shared, fixed initial ($t=0$) network state. The Who to Follow (WTF) algorithm dramatically increases in-degree inequality and clustering producing the visible core-periphery structure, with the high in-degree nodes on the outside, and followers clustered around. This is less pronounced in the local(similar) and bridge(opposite) algorithms. Local$(y)$ algorithms cluster nodes together more tightly than the $\text{bridge}(y)$ algorithms, which have a more diffuse core, for example in bridge(opposite) (B-opp). This allows nodes with contrasting opinions to meet more frequently, and drives cooperation, especially when subject to small network sizes (five initial cooperators). Final network states shown in (b) do not necessarily correspond to network evolutions with default $N$ i.e., $N=800$. For example the special case is shown for B-sim where given $N<150$, the network separates into two opinion clusters, showing strong structural polarization. Otherwise, all results were simulated with default parameters (See \nameref{sec:methods}).}\label{main_comparison}
\end{figure*}

\section{Results}\label{sec:results}

\begin{figure}[!t]
    \centering
    \includegraphics[width=\columnwidth]{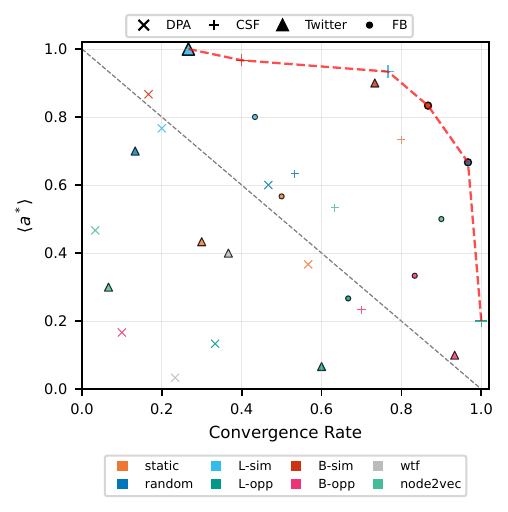}
    \caption{Convergence rates and steady state cooperation, $\langle a^* \rangle$ for all rewiring algorithms under default parameters. The grey reference line $x+y=1$ represents a perfect trade off between convergence rate and final cooperation, i.e., where improvements in convergence rate necessarily reduce final cooperation. Variance is highest in the mid point along the reference and collapses toward a convergence rate of $1$. The red dashed line estimates the Pareto front by indicating scenarios that have no better alternative in terms of final cooperation and convergence rate. Random rewiring and static scenarios represent balanced solutions, but are outperformed by bridge(similar) on the Twitter and FB topologies, which represent the most optimal solutions. Local(similar) on CSF, and random on FB were also optimal solutions, but skewed to favour final cooperation and convergence rate respectively.}
    \label{convergence}
\end{figure}

We study how rewiring algorithms affect depolarization and consensus time by examining both transient dynamics and equilibria of cooperation (mean agent opinion, $a = \sum_{i}^{N}a_i/N$) and polarization (standard deviation of opinion, $P = \sigma(a_1, ..., a_N)$). Here, $N$ is the number of agents in the network. Results are ensemble averages over 90 independent simulation runs, denoted $\langle a \rangle$ for cooperation, and $\langle P \rangle$ for ensemble polarization. Agents change their opinion through pairwise interaction according to Eq. \ref{eq:interaction}. If their opinions differ, interacting reduces the difference unless one agent is a diverger, in which case the backfire effect increases the difference. We also include an agent stubbornness parameter which dampens the effect of  interactions. After interacting, agents' links are rewired according to an algorithm selected from Fig. \ref{algos}. As social change depends more on achieving a cooperative majority than converting all holdouts \citep{andersson_dynamics_2021,otto_social_2020,gardiner_perfect_2011}, we examine the convergence rate within the exponential stage of cluster dissolution, as well as steady state cooperation, $\langle a^* \rangle$, and polarization $\langle P^* \rangle$ . Convergence rate denotes the discrete-time change in cooperation, $\Delta\langle a\rangle/\Delta t$, during depolarization (Fig. \ref{convergence}). The notation $x(y)$ denotes our heuristic algorithms, where $x \in \{\text{local}, \text{bridge}\}$ specifies the topological constraint, and $y \in \{\text{similar}, \text{opposite}\}$ the opinion constraint (similar=homophilic, opposite=heterophilic). 

We analyse how sensitive the equilibrium cooperation values are to the parameters by studying their distributions over 30 independent simulation runs per parameter combination (see \ref{sec:robustness}). Sensitivity to parameter $\theta$ is defined as $S_{\theta} = \sigma(\{A(\theta_i)\}_{i=1}^n)$, where $\theta_i$ represents the $i$-th of $n$ sampled values of parameter $\theta_i$, and $A(\theta_i)$ is the corresponding mean equilibrium cooperation \citep{sobol_global_2001}; lower $S_\theta$ indicates less sensitivity. A full account of the model set up, measures, and parameters is provided in \textit{Methods}. We established the characteristics of each rewiring algorithm group (recommender and heuristic), and their effects. First, relative to static and random baselines (sections \ref{sec:recommenders} and \ref{sec:heuristic}), and then relative to each-other (section \ref{sec:comparing}). Last, in section \ref{sec:robustness}, we show that our findings also hold over a wider parameter space.

\subsection{Representative Dynamics}\label{sec:rep_dynamics}

Figure \ref{main_comparison} shows that the qualitative dynamic stages observed by Andersson et al. for static networks \citep{andersson_dynamics_2021} are maintained in dynamic networks. In the first stage, the slight defector majority at $t=0$ dominates and cooperation declines. Agents with strong opinions spread them to their neighbours, resulting in growing opinion clusters that gradually polarize the network. Interactions within the opinion clusters reinforce polarization. Even without bounded confidence \cite{bernardo_bounded_2024}, this cluster formation stage ends in a fully polarized, homophilous state.
 
 The next stage of the dynamics is a slower process of cluster dissolution (depolarization). Two substages comprise the depolarization stage: First, cooperation grows approximately exponentially. Second, the convergence slows due to shrinking cluster sizes and fluctuations \citep{andersson_dynamics_2021}. In this slower stage of cluster dissolution, meaningful interactions occur only at the cluster perimeters, between agents holding opposing opinions. In this dynamical stage, a slightly positive external field ($\phi$) representing a socio-political environment that encourages cooperation is pivotal. Although this field enables positive steady states ($\langle a^* \rangle > 0$) in all scenarios except WTF, polarization remains high ($0.84$) unless the fraction of diverging nodes ($\rho$) approaches $0$ (see section \ref{sec:robustness}).
 
\subsection{Recommender algorithms}\label{sec:recommenders}

N2V produced higher steady-state cooperation than the static baseline on directed preferential attachment (DPA) networks, while WTF diverged substantially to below this baseline (Fig. \ref{main_comparison}a). Both algorithms underperformed the static baseline on Twitter and the random baseline on both directed networks (Fig. \ref{main_comparison}a,b). Polarization varied between the two algorithms. N2V matched or slightly exceeded the static baseline, whereas WTF reduced steady-state polarization by $0.20$ below the static and $0.30$ below random baselines. This was exaggerated on DPA networks, where WTF had simultaneously the second lowest steady-state polarization and lowest steady-state cooperation. Both N2V and WTF depend on topological structure but interact with it differently. N2V's results align with literature showing that rewiring between structurally similar nodes leads to polarization \cite{santos_link_2021}. It's embedding-based mechanism preserves local network structure while gradually increasing clustering. This emergence of well defined communities (visible in Fig. \ref{main_comparison}b) prevents interactions between different opinions except at community boundaries, which leads to coexistence of low (Twitter, $\langle a^*\rangle \approx 0.24$) to moderate (DPA, $\langle a^* \rangle \approx 0.45$ ) cooperation and moderate (DPA) to high (Twitter) polarization in the steady state.

\begin{figure*}[!t]
    \centering
    \includegraphics[width=\textwidth]{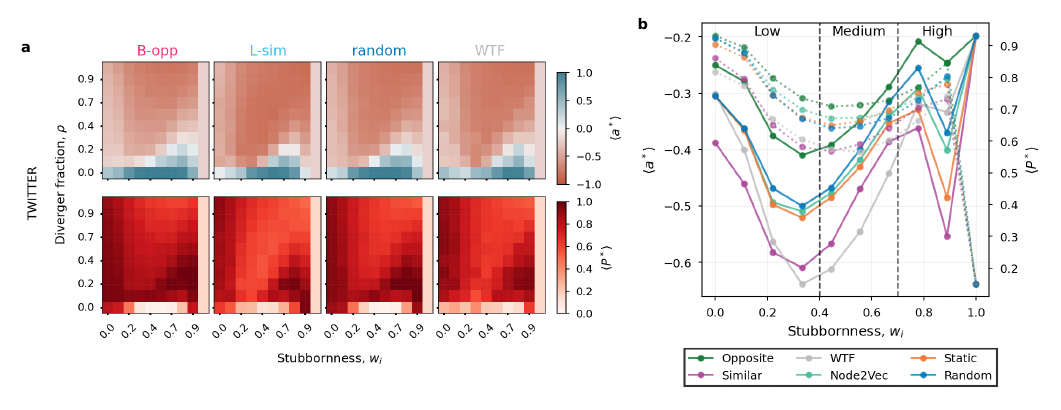}
    \caption{(a) Effect of varying stubbornness, $w$, and diverger fraction, $\rho$ on equilibrium cooperation, $\langle a^* \rangle$ and polarization $\langle P^* \rangle$. Results on the twitter topology are qualitatively similar to other topologies. Cooperation is impossible in the region $\rho > 0.33$ for all algorithm-topology pairs except Who to Follow (WTF) on the directed preferential attachment (DPA) topology (not shown) ($\rho > 0.53$).  Our $x(\text{opposite})$ algorithms (heterophilic) resulted in a greater number of cooperative steady states ($\langle a^* \rangle > 0$), and higher ensemble final cooperation, but also higher polarization. Stubbornness mitigated the otherwise strictly negative effect of divergers on cooperation. (b) Change in final cooperation and polarization over stubbornness values ($w$). Results are grouped by opinion constraint, i.e., opposite and similar.  WTF and $x(\text{similar})$ algorithms produces severely non-monotonic outcomes with increasing stubbornness values. WTF  leads to poor cooperation in the low ($\rho \leq 0.33$) and medium ($\rho \leq 0.66$) stubbornness regimes, which represent plausible interaction conditions in online social media exchanges.}
    \label{heatmap}
\end{figure*}

WTF's interaction with DPA topologies leads to system defection ($\langle a \rangle \approx -0.19$) through a process that reinforces defective clusters in the following way. Visible in Fig. \ref{main_comparison}b, WTF dramatically increases clustering coefficient (from $\approx 0.30$ to $\approx 0.90$) and modularity $Q$ (from $\approx 0.20$ to $\approx 0.80$). This structural reinforcement amplifies existing opinion distributions within tightly clustered communities that predominantly contain (initially) defective agents. Then, the circle of trust mechanism (see \ref{sec:methods})  reinforces connections within topological clusters that converge with defective opinion clusters during early dynamics, and creates echo chambers that lock the system into poor cooperation. This paradoxically reduces polarization ($\approx 0.20$) via homogenization of defective states rather than diverse opinion exchange. On Twitter, higher initial modularity and assortativity produce the opposite behaviour: WTF initially preserves community structure but eventually reduces modularity from 0.70 to 0.30, allowing cooperative pockets to spread and producing moderate cooperation ($\langle  a \rangle  \approx0.31$). Despite this, it creates severe network inequality, increasing in-degree Gini coefficient from $0.50$ to almost $1.0$, which concentrates influence among a few highly connected nodes. 

Our findings are consistent with previous findings \cite{ferrara_link_2022}: on directed networks, N2V moderately increased clustering which is required for spreading processes \citep{everall_pareto_2025}, and reduced cumulative advantage (in-degree Gini), but avoided extreme clustering which can produce echo chambers. On the other hand, WTF dramatically increased both. The observed discrepancy in N2V convergence speed and steady-state cooperation, which both increased by $24\%$ from DPA ($Q=0.20$) to Twitter ($Q=0.62$) is likely due to N2V increasing modularity in comparison to WTF. This allows cooperative pockets to consolidate before spreading. N2V also showed greater robustness to topological changes, contrasting WTF's topology-sensitivity seen here and in other research \cite{espin-noboa_inequality_2022}. Time evolutions of key network metrics are presented in SI Fig. 1.

\subsection{Heuristic algorithms and their variants}\label{sec:heuristic}
Figure \ref{main_comparison} shows that opinion-constrained algorithms (similar/opposite) result in more similar cooperation trajectories than topologically-constrained variants (bridge/local). The $x(\text{similar})$ algorithms also produce higher steady state cooperation than the static baseline ($\langle a^* \rangle  = 0.75$ vs $0.46$), whereas $x(\text{opposite})$ algorithms yield lower cooperation ($\langle a^* \rangle = 0.17$), but faster convergence, particularly in the earlier dynamical phase. These patterns also hold relative to the random rewiring baseline. At low diverger fractions ($\rho = 0.10$),  $x(\text{opposite})$ algorithms force exchanges between agents of large opinion differences ($\lvert a_i - a_j \rvert$), and the positive external field ($\phi$) quickly drives the network to convergence. Under these conditions, regular interactions between disagreeing agents dominate less frequent backfiring interactions.

Whereas cooperation increases for all heuristic algorithms in the first stage of depolarization, dynamics shift in the second stage of depolarization. After $t \approx 3 \times10^4$, the $x(\text{opposite})$ trajectories enter a slow decline, in contrast to $x(\text{similar})$, where cooperation slowly inclines to equilibrium. Homophilic variants achieve higher steady-state cooperation by isolating divergers within homophilic clusters, preventing cascading changes of opinion reversal initiated by divergers. Conversely, heterophilic variants trigger this process, particularly in the second dynamical stage, when exchanges occur at cluster edges between opposing agents. Relative performance of the local and bridge variants depends strongly on the diverger fraction. Without divergers ($\rho = 0$), heterophilic variants $x(\text{opposite})$ achieve faster convergence with steady-state cooperation identical to homophilic variants $x(\text{similar})$ (see \ref{sec:robustness}). Notably, $x(\text{opposite})$ algorithms reach cooperative steady states ($\langle a \rangle \ > 0$) approximately 21\% faster than $x(\text{similar})$ algorithms.

\begin{table*}[t]
\centering
\caption{Algorithm performance vs random rewiring (baseline) taking into account results from default parameters as well as sensitivity analyses. Results for heuristic algorithms varied principally with the opinion constraint, i.e., opposite vs similar, and we group them here respectively.}
\label{tab:algorithm_comparison}
\small
\begin{tabular}{lp{2.5cm}p{1.8cm}p{1.7cm}p{1.6cm}}
\toprule
\textbf{Algorithm} & \textbf{Use-case}& \textbf{Cooperation, $\langle a^* \rangle $}& \textbf{Polarization, $\langle P^* \rangle $}& \textbf{Convergence rate}\\
\midrule
\textbf{Heterophilic: $x(\text{opposite})$}&
Urgent consensus; weak/neutral environment; medium-high stubbornness.&
Higher &
Lower &
Faster \\
\midrule
\textbf{Homophilic: $x(\text{similar})$}&
Consensus focus; favorable socio-political environment; low divergers; medium stubbornness.&
Higher at default params, lower overall&
Much lower &
Slower \\
\midrule
\textbf{WTF (Who to Follow)}&
High stubbornness; in specific cases, with unfavourable socio-political environments.&
Much lower &
Lower &
Slower \\
\midrule
\textbf{N2V (node2vec)} &
Uncertain conditions; preserve structure &
Lower &
Slightly lower &
Comparable \\
\midrule
\textbf{Random} &
High uncertainty; neutral environment; balanced needs.&
Reference &
Reference &
Reference \\
\bottomrule
\end{tabular}
\vspace{2mm}
\par\noindent\footnotesize{Note: Full sweeps $\rho \in [0, 1]$, $w \in [0, 1]$, $\phi \in [-0.05, 0.10]$}
\end{table*}

\subsection{Comparing heuristic vs recommender algorithms}\label{sec:comparing}
Figure \ref{convergence} reveals a trade-off between convergence speed and steady-state cooperation across algorithms and topologies. However, several algorithms achieve Pareto-optimal solutions (red dashed line) that outperform this constraint. We identify Pareto-optimal algorithms as those for which no alternative algorithm achieves both higher convergence speed and higher steady steady-state cooperation. Bridge(similar) dominates on Twitter and Facebook topologies, achieving both rapid convergence ($>0.8$), and high cooperation ($>0.8$), while random rewiring emerges as a balanced generalist solution. The $x(\text{opposite})$ algorithms show comparably fast convergence on directed topologies while maintaining positive (albeit lower) cooperation. In contrast,  WTF shows extreme variability on DPA (cooperation range: $-0.58 < \langle a^* \rangle < 0.29$).  Moreover, WTF shows slower convergence than $83\%$ of other scenarios and produced the lowest cooperation among all tested algorithms.  Although the magnitude of WTF-driven depolarization on the DPA topologies was comparable to $x(\text{similar})$ algorithms, this resulted from the undesirable spread of defection rather than cooperation. The other recommender, N2V, performed consistently across topologies but failed to achieve Pareto optimality, i.e., showing either inferior convergence speed (vs $x(\text{bridge})$) or steady-state cooperation (vs $x(\text{similar})$).

Examining steady-state cooperation outcomes more closely, heuristic algorithms generally outperform recommenders. Patterns of advantage were consistent: $x(\text{similar})$ variants produced higher steady-state cooperation ($0.75$ vs $0.25$), and maintained lower steady-state polarization ($\Delta$-$0.16$) compared to N2V and WTF.  A cooperation level of $0.75$ suggests homophilic rewiring can approach full cooperation under a favourable socio-political environment ($\phi)$, which is notably high for a polarized network model without explicit coordination mechanisms such as optimised interaction scheduling or adaptive rewiring rates \cite{meng_dynamics_2024}. The advantage of homophilic rewiring intensifies on directed topologies, where it does not display the extent of sensitive dependence on network structure seen in N2V and WTF results (section \ref{sec:recommenders}). Comparing local and bridge variants of $x$(similar) across DPA and Twitter topologies yields a difference of $2.5\%$ in steady state cooperation, in contrast to $8-20\times$ larger differences for recommender algorithms (N2V: $0.21$, WTF: $0.50$). A smaller difference is observed for steady-state polarization ($<2\%$ for both DPA and Twitter). 

Two additional elements stand out in the broader comparison: (1) invariance of steady-state cooperation to topological constraints ($<3$\% difference) despite bridge and local algorithms connecting nodes in very different ways. Opinion constraints dominate topological constants for cooperation outcomes. (2) $20 \times$ higher sensitivity to network structure for recommender algorithms shows those designed to be opinion-agnostic can still produce highly context-dependent outcomes, which could explain differential algorithmic amplification across social media platforms (e.g Reddit vs X/Twitter).

Overall, relative to random rewiring, only one of eight algorithms converged faster, and only half achieved higher steady-state cooperation; WTF and N2V were in the lower half for both. Furthermore, random rewiring showed consistent results across topologies, which, coupled with it's ability to assist steady-state cooperation and convergence rate, highlights its versatility. Finally, directed networks produced lower final cooperation ($\langle  a ^* \rangle=0.36$ vs $0.48$) and convergence rates (Fig. \ref{convergence}) compared to undirected networks. This is consistent with prior findings, that in-degree inequalities limit opinion exchange \cite{karimi_homophily_2018}. This has direct implications for platform design choices between reciprocal (Facebook–style) and asymmetric (Bluesky or X) following relationships.

\subsection{Algorithmic robustness}\label{sec:robustness}

Divergers, agents who show the backfire effect in opinion interactions (see \ref{sec:methods}: Eq. 5), impact dynamics over a range of plausible diverger fractions $\rho \in [0.10-0.80]$ (Fig. \ref{heatmap}a). Notably, final cooperation declines monotonically with increasing $\rho$ across algorithms. Also visible in panel Fig. \ref{heatmap}a is the transition from cooperative to uncooperative steady states along the boundary running diagonally from $\rho \approx 0.05$ at low stubbornness to $\rho \approx 0.2$ at high stubbornness. Diverger fractions above this prevent cooperative steady states, as backfiring interactions dominate dynamics. An exception was observed for WTF (on DPA). Overall, the $x(\text{opposite})$ algorithms produce the highest final cooperation at each $\rho$ value (SI Fig. S2). Although this may seem surprising as the $x$(similar) algorithms explicitly avoid backfiring interactions, system cooperation under these scenarios only shows marginally lower sensitivity ($S_{\rho,x(\text{similar)}}=0.35$ vs $S_{\rho,x(\text{opposite})}$) to diverger fraction. The faster opinion exchange facilitated by the $x(\text{opposite})$ algorithm under a positive external field ($\phi$) dominated dynamics even at higher  $\rho$. Stubbornness also mitigated these backfire effects. As stubbornness increased, so did the frontier of cooperative steady states along the $\rho$ axis in Fig \ref{heatmap}, because stubbornness dampened the effect of agent interactions, allowing the (positive) external field to dominate dynamics. Unlike the other algorithms, $x(\text{opposite})$ and WTF algorithms allowed cooperative steady states even at very high stubbornness ($w> 0.80$), extending the cooperative steady-state regime in the case of WTF (on DPA) to $\rho = 0.56$.

To highlight where algorithms excel, we separated final cooperation and polarization results into three regimes (low, medium, high) per parameter. Figure \ref{heatmap}b shows results for stubbornness, on which steady-state cooperation depends most sensitively and non-monotonically. We group $x(\text{opposite})$ and $x(\text{similar})$ algorithms as results vary less with topology (local vs bridge) than opinion constraint (similar vs opposite). Under default parameters (\ref{sec:heuristic}), the $x(\text{similar})$ algorithms excel but, across the $(\rho, w)$ parameter space, $x(\text{opposite})$ variations were generally more insensitive to parameters (cooperation: $\langle a \rangle \approx -0.30$ vs $-0.45$; stubbornness sensitivity: $S_\rho=0.07$ vs $0.12$). However, $x(\text{opposite})$ variants produced higher average polarization ($0.72$ vs $0.63$) and roughly double the frequency of high polarization steady-states ($\langle P ^*\rangle\geq 0.8$). This aligns with prior work showing that heterophilic rewiring, in contrast to homophilic rewiring, better achieves consensus in polarized networks  \cite{borges_how_2024}.

WTF shows low final cooperation in low-medium stubbornness regimes but improves at high stubbornness (Figs. \ref{heatmap}). As discussed (section \ref{sec:recommenders}), WTF creates severe structural inequality (in-degree Gini $\to 1$), concentrating influence in hub nodes, and high clustering ($C\approx0.9$, initial $\approx 0.2$). In high-stubbornness regimes, hub nodes act as stable opinion anchors. They propagate opinions through tightly-clustered communities via the circle of trust mechanism. High modularity ($Q \approx0.8$, initial $\approx 0.2$ ) isolates divergers, preventing opinion reversal cascades. Hub nodes drift toward cooperation with the external field, and tightly-connected clusters follow in a ratchet effect, creating high opinion inertia rather than late-stage oscillations. Exactly this hub-driven effect has been reported under low-frequency (agent) update rate regimes in other work  \citep{meng_dynamics_2024}.  However, this counter-intuitive performance only emerges in a limited parameter space where high stubbornness prevents the formation of a defection trap (see \ref{sec:recommenders}).

Beyond agent-level heterogeneity, we examined algorithmic robustness to the external field (SI Fig. S3). Without an external field ($\phi = 0$), steady state cooperation in most scenarios is negative. Exceptions were $x(\text{opposite})$, random, and WTF. Most scenarios undergo a regime shift at $\phi \approx 0.02$, converging to $\langle a^*\rangle = 1$, while WTF and, to a lesser extent, local(similar) on DPA, show multistability across $\phi$ values. WTF shows strict bistability on DPA. At fixed $\rho = 0.10$, $x$(opposite) variants perform best across topologies, achieving highest cooperative state fractions ($\langle a^*\rangle > 0$): $0.77$ for local(opposite) and $0.75$ for bridge(opposite). Once again, heterophilic rewiring creates conditions for (positive) cooperation with clearer transition boundaries. Directed topologies (Twitter, DPA) show more disorganized, asymmetrical opinion distributions, reflecting in-degree inequality effects.

\section{Discussion}\label{sec:discussion}

Our results show that, compared to a static network, dynamic network structure can lead to the desired outcomes of faster convergence, higher steady-state cooperation, or lower steady-state polarization. However, these outcomes seldom co-occur. The presence of divergers  ($\rho > 0$) will lead to more complex dynamics. Algorithms that converge fastest may lead to higher final polarization, or lower final cooperation (e.g. N2V and WTF). We saw faster convergence and higher steady-state cooperation when links are established to agents holding different opinions, or if they are selected at random. If instead agents establish links homophilically, convergence is slower than on a static network and cooperation does not necessarily emerge in the wider $(\rho, w)$ parameter space. However, steady-state polarization is also low in such cases. Essentially, performance depends on regime and context. This raises the question of desired outcome. Does the system need urgent change or high long-term cooperation? We summarise where each algorithm performs best in Table \ref{tab:algorithm_comparison}. Despite the complexity, some consistent trends existed across algorithms. Final cooperation increased with stubbornness up to a high stubbornness regime $w > 0.8$. An increase in diverger fraction reduced final cooperation but stubbornness mitigated this effect. This was particularly visible for the WTF and $x(\text{opposite})$ algorithms, which reached cooperative system steady states with diverger-dominated populations. Further, cooperative outcomes are assisted by a strong pro-cooperation sociopolitical environment but can also emerge in its absence under certain conditions. We saw this with heterophilic opinion constraints, i.e., $x(\text{opposite})$, and under random rewiring.

Based on our results, WTF algorithms fail to produce steady-state cooperation under standard parameters and also low to mid level stubbornness and diverger fractions. Our findings are consistent with other research showing that WTF can increase polarization and decrease other network properties conducive to collective action \cite{su_effect_2016, ferrara_link_2022, espin-noboa_inequality_2022}. For example, we corroborate the finding that WTF tends to raise the in-degree Gini coefficient, which hampered interactions between differently behaving agents in our model. It was also the drastic increase in clustering, combined with DPA's topological starting conditions which locked in a defective system on the DPA network under default parameters. Counterintuitively, an algorithm designed to connect users produces system-wide defection given certain network structures. It highlights that even opinion-agnostic algorithms can produce strongly different outcomes in social dynamics across different network structures.

One positive exception is the ability of WTF to sustain cooperative opinion change under high stubbornness, which dampens interactions.  This result is broadly consistent with prior work \citep{meng_dynamics_2024} in opinion dynamics on networks with hubs (high-degree nodes).  If hubs can be somehow locked in as cooperators, a planner could utilise the hub-follower network structure formed by the WTF algorithm to guide the system to cooperation. Such a method could be useful in some situations where the type of behaviour is recalcitrant or difficult to change, as with dietary behaviour or religious belief \citep{everall_pareto_2025}.  N2V was less sensitive to topological changes than WTF.  This may be related to the mechanism of N2V recommendations which rely on embedding-space proximity, not necessarily topological or opinion closeness. Prior work has also found that N2V helps mitigate rich-get-richer network effects \cite{ferrara_link_2022}. Embedding-space proximity or similar techniques could thus play a role in generating, or preserving network structures that hinder cascading effects which push the system into undesirable states, such as low final cooperation.

On the question of homophily vs heterophily, our finding, that homophilic rewiring is less able to drive cooperation in more polarised system states, is consistent with those by Axelrod et al. \citep{axelrod_preventing_2021} and Nguyen \citep{nguyen_echo_2020}, that interactions between polarized clusters can act to maintain polarization when opposite opinions are actively discredited through manipulations of trust. To dissolve such echo chambers, access to agents and information sources outside echo chambers need to be enabled \citep{nguyen_echo_2020, borges_how_2024}. We observed this, where the $x(\text{opposite})$ algorithms produced low but positive steady-state cooperation under default parameters, and the overall highest steady-state cooperation of all algorithms over the $(\rho, w)$ parameter space. However, we also found that heterophilic rewiring was effective at higher diverger fractions ($\rho$), when there were more backfiring interactions and agents were more stubborn.

Furthermore, the finding that cooperation under $x(\text{opposite})$ algorithms was robust to higher diverger fractions at middle and higher stubbornness values suggests that heterophilic algorithms can be useful to promote consensus even in fractious social contexts. These more stubborn interactions are also more similar to those under complex contagion dynamics, in which agents require multiple exposures to a different opinion to change their opinion. Existing work \cite{borges_how_2024} parallels our findings, showing that under complex contagion, heterophilic rewiring plays a key role in facilitating consensus. The authors add that under simple contagion, or with more labile agents, homophilic rewiring is more suitable. However, even categorized into the dichotomy of simple or complex contagion, it is difficult to determine which applies in any given social context \citep{fink_investigating_2016, horsevad_transition_2022}. Others suggest simple contagion-type exchanges are a general feature of modern digital interaction \cite{bak-coleman_stewardship_2021}. The continuing uncertainties  around the dominant dynamic in different social interactions, i.e, whether they are more similar to complex or simple contagion spread, preclude prescribing heterophilic or homophilic rewiring regimes as a single solution to fostering cooperation. 

Our study combined many features of real-world social network dynamics to tease apart how their complex interplay affects prosocial consensus building. To maintain interpretability, this required numerous simplifying assumptions that also limit the generality of our results.  Opinion constraints in our heuristic rewiring algorithms are limited to a binary decision (homo- or heterophily). In reality, interaction likelihoods depend on a continuous distribution of preferences. Similarly, agent stubbornness is fixed and homogenous. Varying the stubbornness between agents would imply coexistence of easily persuaded and intransigent agents. We might expect the former to cluster around the latter, who are effectively `influencers', thus reinforcing polarization \cite{andersson_dynamics_2021}. Multidimensional vectors can substantially affect opinion dynamics \cite{baumannEmergencePolarizedIdeological2021}.  Further, expanding the current one-dimensional agent opinion state into a multidimensional vector would represent multiple character and cultural traits \citep{Pham2021, Axelrod1997}. Such an approach could analyse how social media algorithm design functions when commonalities, rather than differences are highlighted\citep{dunbar_anatomy_2018, gladston_social_2019}. 
 
A more complex social interaction model could be realised with a generative ABM \cite{vezhnevets_generative_2023} (GABM,) where large language models (LLMs) prompted with social media data such as user profiles constitute the agents. Along with the existing context provided by the LLM training data, this would allow a richer representation of agents with multidimensional traits. Recent work applying GABMS in the context of news sources \cite{tornberg_simulating_2023} showed that opinion-heterophilic news article recommender algorithms were effective at building consensus, similar to our bridging algorithm.  However, even with the added complexity provided by GABMS, equation based social interaction models are still able to recreate similar dynamics. This is shown in recent work focusing on echo chamber formation in online communities. An equation-based social interaction term similar to that used in our model performed reasonably well compared to a GABM approach featuring six different LLMs when validated against empirical datasets \cite{gu_large_2025}. Finally, we assumed that newly established connections are stable whether or not they connect agents holding similar or opposite opinions. This is contrary to evidence that disagreeing individuals tend to avoid contact \citep{dunbar_anatomy_2018}, and that connections are dynamic, typically decaying over time \cite{roberts_costs_2011}. Future extensions of this work could address this by adding link decay rates that depend on the agents' states.

Our results show that our simple heuristic algorithm which joins people of different opinions vastly outperformed a WTF–based rewiring algorithm in enabling lasting social cooperation. Moreover, even our random rewiring algorithm similarly outperformed WTF. The latter finding supports prior indictments of recommender algorithms  \cite{espin-noboa_inequality_2022, lasser_designing_2025}. To reduce time-to-consensus (by approximately 21\%) and promote cooperation it is thus advisable to connect people holding different opinions. This work provides theoretical support to the substantial body of evidence that polarization in social media could be limited or reversed if recommender algorithms were modified to promote and maintain links between groups holding opposite opinions (i.e out-groups) \citep{santos_link_2021, pappalardo_survey_2024}. This holds even in the face of the backfire effect, or a less agreeable external field. When little is known about a given social network, randomly rewiring links between people can be an effective tool to encourage cooperation and reduce polarization.

\section{Methods}\label{sec:methods}

\begin{table*}[t]
    \centering
    \caption{Summary of default model parameters.}
    \label{tab:parameters}
    \begin{tabular}{lp{5cm}l}
        \toprule
        \textbf{Parameter/Variable}          & \textbf{Represents}                                                                            & \textbf{Value} \\
        \midrule
        External field (\(\phi\))       & Global factors: Policy regulation, mass media, morals, etc                            & 0.05 \\
        Stubborness (\(w_i\))& Resistance to influence: Stubbornness, ideological commitment, etc                    & 0.6 \\
        Agent-agent weight (\(w_{ij}\)) & Susceptibility: Friendship, animosity, charisma, etc                                  & 0.5 \\
        Agent-agent weight SD     & Susceptibility variance                                                               & 0.15 \\
        Initial cooperation    & Initial average cooperation                                                         & -0.25 \\
        Initial cooperation SD      & Variance in initial cooperation                                                     & 0.15 \\
        Randomness (\(r\))          & Noise intensity in agent interactions                                                 & 0.1 \\
        Average degree ($\langle k \rangle$)             & Degree of nodes added during network growth  & 8 \\
        Clustering                  & Probability of a triad formation step in network generation     & 0.5 \\
        diverger fraction ($\rho$)       & Probability of an agent to be a diverger & 0.10 \\
        \bottomrule
    \end{tabular}
\end{table*}
\subsection{Agent-based model}\label{sec:abm}

Each agent's level of behavioural cooperation is represented by a continuous real-valued variable \(a \in [-1,1]\) and is also referred to as the agent's `opinion' since our formalism borrows from opinion dynamics modelling. However, despite this terminology, we stress that each agent's state variable is intended to represent behaviour rather than just opinion. We refer to agents for whom $-1\leq a<0$ as defectors and $0<a\leq1$ as cooperators. We distinguish between topological clusters, detected using the Louvain algorithm \citep{blondel_fast_2008}, and opinion clusters, which are made up of agents sharing the same opinion. Following Andersson et al \cite{andersson_dynamics_2021}, simulations begin with mildly defecting agents because we are concerned with building collective action from a state of inaction.

Each model time-step in an independent network simulation constitutes eight operations, these repeat until the network converges to a steady state:

\begin{enumerate}
    \item Randomly select an agent \(i\)
    \item Randomly select a neighbour \(j\) of \(i\)
    \item Perform interaction between agent \(i\) and \(j\) as described in section \ref{sec:agentinteraction}
    \item Update \(i\)'s opinion according to interaction outcome
    \item Select a non-neighbour \(k\) of \(i\) following the chosen rewiring algorithm
    \item  Establish a new link between \(i\) and \(k\) as with probability $p_{join}=0.5$ (section\ref{sec:linkrw})
    \item Randomly select a neighbour \(l\) of \(i\)
    \item  Break a link between \(i\) and \(l\) as described in section \ref{sec:linkrw}
\end{enumerate}

While the probability of being chosen as agent \(i\) is independent of an agent's degree, the likelihood of being selected as a neighbour \(j\) for interaction increases with $j$'s degree. Updating $i$'s behaviour, rather than $j$'s, ensures that agents with more connections are more influential as their opinion states change more slowly over time.

\subsection{Agent interaction}\label{sec:agentinteraction}

The change in the opinion \(a\) of an agent \(i\) when interacting with a neighbouring agent \(j\) is given by

\begin{equation}
\Delta a_i = \lvert a_i - a_j\rvert [w^- (1-a_i)-w^+ (1+a_i)]
\end{equation}

Where \(w^+\) and \(w^-\) are the weights for an agent becoming more or less cooperative respectively. To simplify we set \(w^+ = f(x_{ij},r,\xi\)) and \(w^- = 1- w^+\) and obtain

\begin{equation}\label{eq:interaction}
\Delta a_i = \lvert a_i - a_j \rvert [2f(x_{ij},r,\xi) - 1 - a_i]
\end{equation}

where,

\begin{equation}
    f(x_{ij},r,\xi) =
    \begin{cases}
    0                           & \text{if $x_{ij}+\xi < -r$}\\
    \frac{1}{2r}(x_{ij}+r-\xi)         & \text{if $x_{ij}+\xi \in [-r,r] $}\\
    1                           & \text{if $x_{ij}+\xi > r$}
    \end{cases}
\end{equation}

and,
\begin{equation}
    x_{ij} = w_i a_i + w_{ij} a_j + \phi
\end{equation}

The function \(f(x_{ij},r,\xi)\) captures the response of the affected agent. The parameters \(w_i\) and \(w_{ij}\) represent the self-weight and the weight between agents \(i\) and \(j\). This captures the local mechanisms of stubbornness as well as the susceptibility, which determines the effect of an agent on another. \(\xi\) is a random noise term that is distributed over the interval \(\pm r\) and represents stochastic external factors that may affect the interaction, such as the mood the agent was in. \(\phi\) captures global effects that influence all agents at all times and can be referred to as the external field. This simulates channels shared by all actors such as relevant political policies, media reporting or the moral hegemony of society\citep{andersson_simple_2020}. In addition to the dynamics described above, we follow Santos et al. \cite{santos_link_2021}, and include a second equation to describe the backfire effect. Where interactions lead to diverging opinions. Due to the difficulty in estimating the prevalence of the diverger effect in a given social interaction, we set our default diverger fraction quite conservatively, but deal with model sensitivity to this parameter in \ref{sec:robustness}. Adapting a similar approach to Santos et al. \cite{santos_link_2021}, we define an alternative version of Eq. \ref{eq:interaction}:

\begin{equation}
\Delta a_i = \lvert a_i - a_j \rvert \left[ 2f(x_{ij}, r, \xi) - 1 \right] [ a_i a_j ]^\lambda
\end{equation}

where \(a_i a_j = -1\) if individuals have opposing viewpoints and \(\lambda \in \{0,1\}\) dictates whether the interaction is governed by a converging dynamic ($\lambda = 0$), or a backfiring dynamic ($\lambda = 1$). In the former we simply obtain Eq. \ref{eq:interaction}. The parameters are chosen according to Andersson et al. \citep{andersson_dynamics_2021} and are summarized in Table \ref{tab:parameters}.

The external field is taken to be \(\phi = 0.05\). It is chosen to be positive in order to reflect a society that incentivizes cooperative behaviour. The agent-to-agent weights \(w_{ij}\) are drawn from a normal distribution with mean \(\mu = 0.5\) and standard deviation \(\sigma = 0.15\). We choose the self-weight to be \(w_i = 0.6\) as it is assumed that people are more likely to keep their own opinion than change it, but are quite prone to opinion change nevertheless \citep{mallinson_effects_2018}. The random noise term \(\xi\) is uncorrelated with the interaction and hence assumed to be uniformly rather than normally distributed. The width of the noise is set to \(r = 0.1\). For a more detailed discussion of parameters we refer to \citep{andersson_dynamics_2021}.

\subsection{Network typology}\label{sec:network}

Social interactions in our model occur between agents embedded in a network, where they are represented as nodes, and their social contacts as links. We define such a network as a graph $G = (V, E)$ with $V$ the set of nodes $V = \{ v_i, ..., v_n \}$, and $E \subseteq V \times V$ as the set of weighted, directed(undirected) links. All our networks have cardinality $\vert V \vert = N = 800$. We employ both synthetic, i.e generated networks, as well as empirical networks in exploring the evolution of model dynamics. These categories are split again into undirected and directed examples. For our undirected synthetic network, we focus on clustered scale free networks and use the Holme-Kim clustered scale-free network growth algorithm \citep{holme_growing_2002} with an average degree of \(\langle k \rangle = 8\). Directly linked agents are referred to as neighbours or friends interchangeably. For our synthetic directed network we use the DPA (directed preferential attachment) model, an extension of the BA (Barabsi-Albert) model \citep{espin-noboa_inequality_2022}. Networks based on these models are generated with the NetIn package in python \citep{pynetin}(see SI Tab. 1 for parameters). We consider two empirical networks representing two popular digital spaces, Twitter (directed) and Facebook (undirected). In Table \ref{tab:network_properties} we summarize relevant network properties, and define which elements of the real social networks are represented as nodes and links.

\begin{table}[!t]
\centering
\caption{Structural properties of empirical networks used in simulations}
\label{tab:network_properties}
\begin{tabular}{lcc}
\toprule
\textbf{Property} & \textbf{Facebook} & \textbf{Twitter} \\
\midrule
Nodes ($N$) & 786 & 789 \\
Edges ($M$) & 14,024 & 12,110 \\
Average degree ($\langle k \rangle$) & 35.7 & 30.7 \\
Clustering coefficient ($C$) & 0.476 & 0.426 \\
Assortativity ($r$) & 0.33 & 0.17 \\
Modularity ($Q$) & 0.52 & 0.63 \\
Avg. path length ($\langle \ell \rangle$) & 3.04 & 4.33 \\
Small-world coeff. ($\sigma$) & 7.32 & 13.6\\
\bottomrule
\end{tabular}
\vspace{2mm}
\par\noindent\footnotesize{*Networks are undirected for Facebook and directed for Twitter.}
\end{table}

\subsection{Link rewiring}\label{sec:linkrw}

Link rewiring allows for a dynamic network structure where the neighbours of agent \(i\) can change over time.  We introduce six rewiring algorithms in total. One random rewiring algorithm as a a baseline. Two of which are topology-based recommender algorithms, and represent part of real-world strategies employed by digital network operators such as Google and Twitter \cite{gupta_wtf_2013} and detailed in section \ref{sec:recommenderalgs}. The remaining three are heuristic and formulated based on a literature review and introduced in section \ref{sec:heuristicalgs}. The latter represent limiting cases on extreme ends of probability spectra. Their purpose is to give first qualitative insights into how different ways of social restructuring can affect the dissolution of polarization. We introduce two algorithm (sub)variants for the heuristic rewiring algorithms, each differing by only one assumption, to isolate the effects of each algorithm. The Who-to-follow algorithm was only tested on directed networks.

\subsubsection{Recommender Algorithms}\label{sec:recommenderalgs}

\paragraph{Who to follow (WTF)}

Introduced by Twitter (now X) \cite{gupta_wtf_2013}, the Who to Follow (WTF) algorithm seeks to connect a user with similar users. Given a set of users $V$ on a network, it does so by calculating a $\textit{circle of trust}$ for each user $u \in V$. It then ranks each user $u \neq i$ using the $\texttt{SALSA}$ algorithm \cite{lempelSALSAStochasticApproach2001}. The user with the highest WTF score who is not directly connected to $i$, but connected through $i$'s circle of trust will be recommended to $i$. The $WTF$ score for a given user $u$ can be expressed:

\begin{equation}
\text{WTF}(u) = \text{SALSA}(\text{COT}_u)
\end{equation}
where
\begin{equation}
\text{COT}_u = \{v \in V \mid \text{PPR}_u(v) \in \text{top}_k\}
\end{equation}
represents the circle of trust for user $u$, determined by the top $k$ nodes with highest personalized PageRank scores relative to $u$, and $SALSA(COT_u)$ computes authority scores in the bipartite graph formed between nodes in the circle of trust (acting as hubs) and their respective neighbour (acting as authorities).

\paragraph{node2vec (N2V)}

An algorithmic framework for a flexible, and scalable feature learning for nodes in networks \cite{grover_node2vec_2016}. N2V produces a low-level feature representation of nodes as embedded vectors in a vector space. This representation retains critical information about the nodes which can also be extended to their edges. It does this by optimising the likelihood that a network neighbourhood of a node is preserved based on it's feature representation. N2V lends itself to link prediction tasks on social networks, where it scores exceptionally well in comparison to other state of the art techniques \cite{saxena_nodesim_2022}. These include both heuristic predictors such as the common neighbours, and preferential attachment algorithms, as well as other feature learning algorithms \cite{grover_node2vec_2016}. Each node $i \in V$ is rewired based on the cosine similarity of their respective vector representation in the embedding space. For a given pair of embedded vector projections $(\vec{v}_i, \vec{v}_j)$, the cosine similarity is:

\begin{equation}
    \theta(\vec{v}_i, \vec{v}_j) = \frac{\vec{v}_i \cdot \vec{v}_j}{\|\vec{v}_i\| \|\vec{v}_j\|}
\end{equation}

where $\theta$ is the cosine similarity, $\vec{v}_i \cdot \vec{v}_j$ is the dot product of the vectors, and $\|\vec{v}_i\|$, $\|\vec{v}_j\|$ are the norms of the vectors, respectively. The vector projection $\vec{v}_j$ which maximises \(\theta(\vec{v}_i, \vec{v}_j)\) with respect to a given \(\vec{v}_i\) will be recommended to $i$. More information on N2V and a reference implementation can be found on github \cite{groverAdityagroverNode2vec2025a}. For our model, we use a SNAP implementation in C++ \cite{snapSnapstanfordSnap2025}.

\subsubsection{Heuristic Algorithms}\label{sec:heuristicalgs}

\paragraph{Random rewiring}

A baseline scenario in order to evaluate the role of unguided network change. Agent \(i\) establishes a link to a random non-neighbour \(k\) with a probability \(p\). With a probability \(q\) a link between agent \(i\), and a neighbour, \(l\) is removed. \(q\) does not change with the number of neighbours. The finding that cognitive capacity and available time limit the number of friends an individual is able to hold \citep{dunbar_anatomy_2018} is included only indirectly, as an agent with high degree has a higher likelihood of being chosen as the neighbour with whom a link is cut. To maintain a stable average degree we assume \(q=2p\).

\paragraph{Local rewiring}

The local rewiring algorithm includes two tendencies found in empirical studies to simulate simplified, but realistic friendship establishment. Firstly, an individual's social surroundings make them more likely to engage with people who are topologically close \citep{santos_link_2021,nasiri_new_2021,dunbar_anatomy_2018}. Secondly, individuals' engagements are subject to homophily, meaning they will most likely show preferential attachment to individuals with whom they have a lot in common \citep{williams_network_2015,dunbar_anatomy_2018,kozma_consensus_2008}.

For implementation we use an approach similar to Crawford et al. \citep{crawford_resisting_2018}. Figure \ref{algos}A illustrates our assumptions. The colour of a node depicts its opinion. We focus on agent \(i\) and limit the pool of available agents \(k\) for rewiring to those non-neighbours that are within \(d\) network steps. These are marked with dashed circles. We set \(d = 2\) (i.e. agent \(i\) can only choose among friends of friends). This is a practical choice as \(d = 3\) ($\langle k \rangle ^3 = 512$) would allow agent \(i\) to establish links with almost half of our chosen network. For breaking a link, a random neighbour \(l\) of agent \(i\) is selected. To maintain a constant average degree, a link is broken if and only if a link is established.

\paragraph{Bridge rewiring}

In the bridge rewiring algorithm we incorporate recommendations for dissolving polarization. They focus on inter-community interactions and their importance as information channels to increase individuals' exposure to differing views \citep{andersson_dynamics_2021,gladston_social_2019,williams_network_2015,lee_homophily_2019,fotouhi_conjoining_2018}. As figure \ref{algos}C illustrates, we limit the potential new neighbours of agent \(i\) to the non-neighbours outside of \(i\)'s own cluster. Clusters are identified by the Louvain algorithm \citep{blondel_fast_2008}. Nevertheless it is likely that multiple communities of defectors separated by ones of cooperators exist within a network.

\paragraph{Rewiring variations}

Two additional assumptions are tested in the local and bridge rewiring algorithms with an opinion constraint. A notation of (similar) and (opposite) is added to indicate whether links are established only if agents are of the same opinion (i.e homophily) or only if they are of a different opinion (i.e heterophily). Local(similar) therefore refers to the rewiring algorithm in which agent \(i\) chooses a random agent \(k\) within two network steps and establishes a link if they are of the same general opinion (see Figure \ref{algos}) i.e., if \(a_i \wedge a_k < 0\) or \(a_i \wedge a_k \geq 0\). Local(opposite) describes the inverse, i.e., if \(a_i \geq 0 \wedge a_k < 0\) or \(a_i < 0 \wedge a_k \geq 0\). Bridge(similar) (see figure \ref{algos}D) and bridge(opposite) (see figure \ref{algos}C) both limit the potential neighbours of agent \(i\) to non-neighbours outside of \(i\)'s own cluster. However, a link is established only if the agents agree in the former, and only if they disagree in the latter.

\backmatter

\bmhead{Acknowledgements}

We thank D. Andersson, who provided us with the initial code and significantly helped during the first phase of the project. We also thank Fariba Karimi, who provided valuable feedback on the manuscript.

\section*{Declarations}

\subsection*{Funding}
The authors acknowledge financial support from the University of Graz.

\begin{appendices}

\end{appendices}

\bibliography{sn-bibliography}

\end{document}